\documentclass[onecollarge]{svjour2}       
\smartqed  
\usepackage{graphicx}
\usepackage{mathptmx}      
\usepackage{amssymb}
\usepackage{amsmath}
\newcommand{\vu}{\mathbf u}

\begin{document}

\title{New prospects for de Broglie interferometry}
\subtitle{Grating diffraction in the far-field and Poisson's spot in the near-field} \dedication{This paper is dedicated to Daniel Greenberger and Helmut Rauch, two pioneers in explorations of the {\em foundations of quantum physics}.}

\titlerunning{Limits in matter wave diffraction}
\author{Thomas Juffmann, Stefan Nimmrichter and Markus Arndt  \and Herbert Gleiter \and Klaus Hornberger}

\authorrunning{Juffmann et al.} 

\institute{Thomas Juffmann, Stefan Nimmrichter and Markus Arndt\at
              Faculty of Physics, University of Vienna, Boltzmanngasse 5, 1090 Vienna, Austria \\
              Tel.: +43-1-4277 51210, Fax: +43-1-4277 9512 \email{markus.arndt@univie.ac.at}           
           \and
            Herbert Gleiter \at Institut f\"{u}r Nanotechnologie, Karlsruhe Institute of Technology, Karlsruhe, Postfach 36 40, 76021  Karlsruhe, Germany
            \and
           Klaus Hornberger \at Max Planck Institute for the Physics of Complex Systems, N{\"o}thnitzer Str. 38,    01187 Dresden, Germany }
\date{Received: date / Accepted: date}
\maketitle

\begin{abstract}
We consider various effects that are encountered in matter wave interference experiments with massive nanoparticles. The text-book  example of {\em far-field interference at a grating} is compared with diffraction into the dark field behind an opaque aperture, commonly designated as {\em Poisson's spot} or the {\em spot of Arago}. Our estimates indicate that both phenomena may still be observed in a mass range exceeding present-day experiments by at least two orders of magnitude. They both require, however, the development of sufficiently cold, intense and coherent cluster beams. While the observation of Poisson's spot offers the advantage of non-dispersiveness and a simple distinction between classical and quantum fringes in the absence of particle wall interactions, van der Waals forces may severely limit the distinguishability between genuine quantum wave diffraction and classically explicable spots already for moderately polarizable objects and diffraction elements as thin as 100\,nm.
\end{abstract}


\section{Introduction}
Quantum physics ranks among our best-confirmed concepts of nature. And yet, a conceptual gap in the transition between quantum physics and classical observations has not yet been overcome.
Matter wave interferometry with massive particles~\cite{Broglie1923} has always been paradigmatic for the peculiar predictions of quantum physics as it demonstrates quantum delocalization and the superposition principle for material particles such as free electrons~\cite{Hasselbach2010}, neutrons~\cite{Rauch2000}, atoms~\cite{Cronin2009}, dimers~\cite{Bord'e1994} and complex molecules~\cite{Arndt1999} during their unperturbed propagation.

Much of the pioneering work on matter wave interferometry is in particular associated with developments in neutron quantum optics. And in this issue we celebrate the birthdays and work of two central figures in research on the foundations of quantum physics, Helmut Rauch~\cite{Rauch1974} and Daniel Greenberger~\cite{Greenberger1983}.
A number of recent advances both in the cooling of micromechanical oscillators (see collection in~\cite{Aspelmeyer2008}) and in matter wave manipulation methods~\cite{Arndt1999,Brezger2002,Reiger2006,Gerlich2007,Arndt2009} now promise future extensions of experiments testing the superposition principle to much higher mass and complexity. In this paper we focus on two particularly simple concepts of particle interferometry, namely far-field diffraction behind a grating and near-field interference behind an opaque sphere or disk, i.e. the observation of Poisson's spot. Recent developments of new nanoparticle sources~\cite{Haberland1991,Deachapunya2008,Marksteiner2008} and novel detection methods~\cite{Juffmann2009} may soon allow one to experimentally access them in a mass regime between $10^4$ and $10^6$ atomic mass units (amu).

{\em Grating diffraction} has already been thoroughly studied with electrons~\cite{Jonsson1974,Freimund2001,Gronniger2005}, neutrons~\cite{Gahler1991}, atoms~\cite{Martin1988,Keith1991,Carnal1991} and molecules~\cite{Chapman1995,Schollkopf1996,Nairz2001a,Nairz2003}.
The {\em Poisson spot} was observed with matter waves for the first time with electrons~\cite{Komrska1971,Matteucci1990} and later extended to 1D diffraction behind a wire and 2D interference behind either a free disk or a zone plate using neutrons~\cite{Kearney1980,Gahler1991}, atoms~\cite{Nowak1998,Doak1999} and most recently also the diatomic molecule D$_{2}$~\cite{Reisinger2009}.

\begin{figure}
\centering
  \includegraphics[width=0.8\columnwidth]{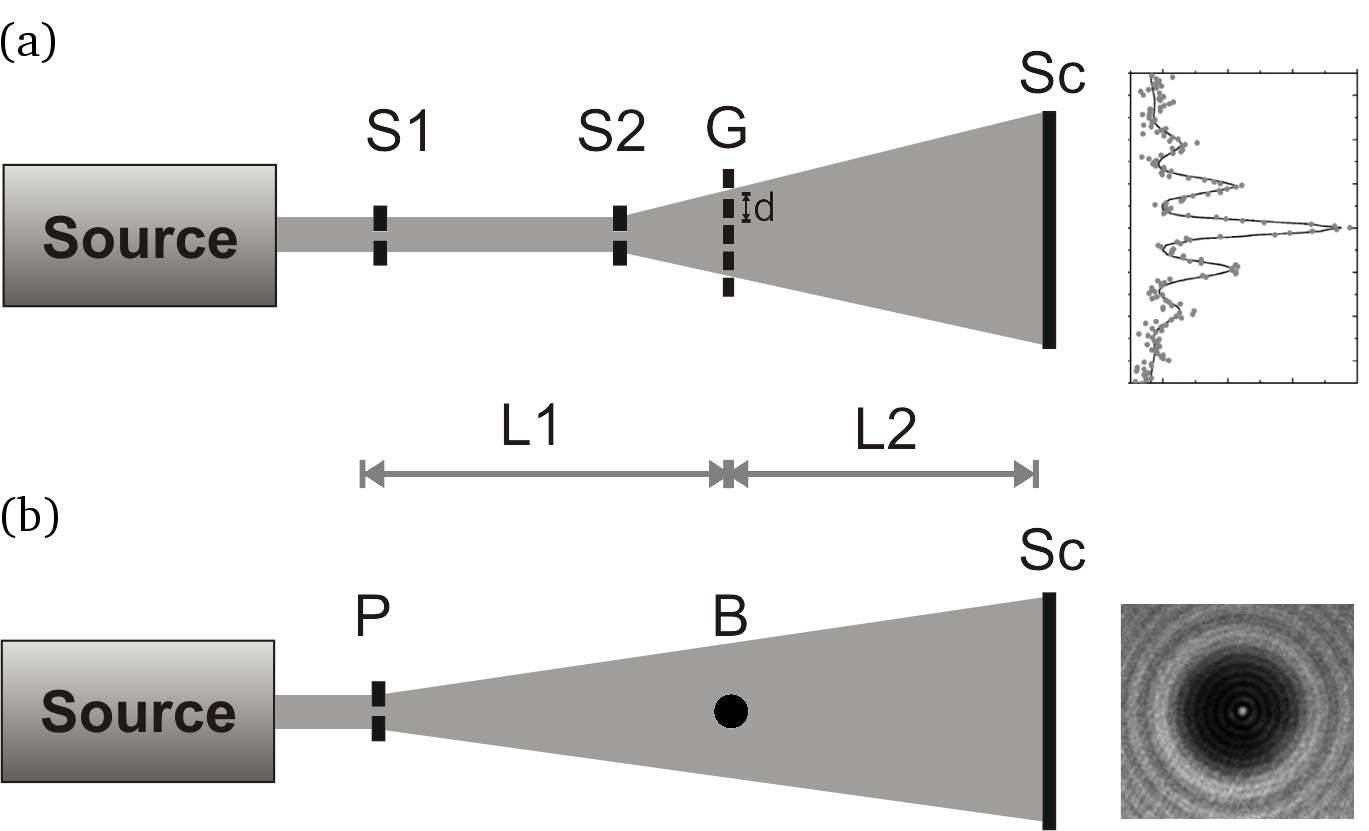}
  \caption{(a) In a far-field diffraction experiment  the collimation slits S1 and S2 prepare the transverse coherence and the collimation required in order to resolve an interference pattern on the screen Sc behind the grating G. The interference pattern in the figure represents experimental data for $C_{60}$ molecules from~\cite{Nairz2003}. (b)
The Poisson spot experiment is based on a
radially symmetric setup consisting of a small pinhole P and an opaque aperture B. Wave diffration at the edge of the circular aperture leads to a bright interference spot located at the center axis in the geometric shadow. The depicted interference pattern is an illustration with laser light that was observed behind a 1\,mm sphere illuminated with at a wave length of 532\,nm.}\label{Fig1}
\end{figure}

In the following we assume quantum physics to be the correct theory for arbitrary particle size and mass, putting aside recently suggested modifications of standard quantum theory ~\cite{Diosi1989,Diosi2004,Penrose1996,Bassi2003,Wang2006,Carlip2008,Adler2009}. Instead, we ask which experimental constraints will in practice be limiting matter wave observations when the complexity of the particles grows.

The de Broglie wavelength $\lambda_\mathrm{dB} = h/mv$ of a particle at speed $v$ determines the size of the diffraction pattern. If the particle source is in contact with a thermal bath the  most probable de Broglie wavelength $\lambda_\mathrm{th} = h/(2kTm)^{1/2}$ will be related
to the most probable particle velocity  $v_\mathrm{mp} = (2k_BT/m)^{1/2}$ where the temperature $T$ is measured in Kelvin and $k_B$ is Boltzmann's constant. Present-day interferometers are designed to deal with matter waves with wave lengths of about $\lambda_\mathrm{dB} \simeq 1$\,pm. This corresponds to a C$_{60}$  fullerene with a mass of $720$\,amu at $v=550$\,m/s  or equivalently to the gold cluster Au$_{5000}$  with a mass of about $ 10^{6}$\,amu and $v=0.4$\,m/s. The latter is close to the thermal velocity at about 10\,K and in reach of cryogenic buffer gas technologies inside a cold ion trap. In the following we compare the diffraction of C$_{60}$ at a most probable velocity of 150\,m/s, corresponding to a thermal beam at 900\,K (case 1), to the case of Au$_{5000}$ at  v=1\,m/s  (case 2) .  The polarizabilities are taken to be $89\, \AA^3$ for C$_{60}$ \cite{Berninger2007} and $2.5\times 10^4\,\AA^3$  for the gold cluster \cite{Wang2006b}.

\section{Far-field matter wave diffraction}

Far-field diffraction as depicted in Fig.~\ref{Fig1}(a) faces several requirements.
First, the particles must be smaller than the grating period in order to neither get stuck in a material mask nor to average over neighboring potential wells in case of an optical diffraction grating.
Second, both the beam diameter and its transverse coherence have to cover at least two slits, separated by the distance $d$. This condition is met if the collimation angle  $\Theta$  of the molecular beam satisfies $\Theta = D/2 L_1 <\lambda_\mathrm{dB}/d$.  We assume a symmetric setup, where the source of width $D$ acts as the first collimator, equal in width to a second collimation slit at a distance $L_1$ further downstream.

\emph{Mechanical nanogratings}, with slit openings as tiny as 50\,nm, periods of 100\,nm are close to the smallest structures that can currently be made. Assuming a typical grating membrane thickness $b\simeq 100$\,nm, the van der Waals interaction between the traversing molecules and the slit wall leads to a significant attractive force  which results in the narrowing of the effective slit width~\cite{Grisenti1999,Nairz2003}. Since the effect of this dispersion force grows with increasing polarizability and decreasing velocity $v$, a particle may  even be adsorbed by the surface if it approaches it within a cutoff distance ~\cite{Nimmrichter2008}
\begin{equation}\label{CutOff}
x_c= (18 C_4 b^2/m v^2)^{1/6}.
\end{equation}
This estimate is based on the asymptotic form of the Casimir-Polder potential, with the constant  $C_4 = 3\hbar c \alpha / 8 \pi$.  In case 1, all fullerenes  that approach the wall within 17\,nm will be removed from the beam. For the Au$_{5000}$ cluster (case 2) the cutoff distance amounts to already 46\,nm. This reduction of the useful slit width indicates that there is a technical limit for grating diffraction. Ultra-thin membranes -- made for instance from atomically thin graphene~\cite{Geim2007} or nanometer-sized graphenoids~\cite{Schnietz2009} -- appear therefore very promising for future diffraction experiments if they are to be performed with material gratings. It is yet still necessary to show that such nanosheets can be prepared with the required accuracy and mechanical stability.

 \emph{Optical absorption}~\cite{Reiger2006}  or \emph{phase gratings}~\cite{Nairz2001a}
do not suffer from this limitation, but they are also intrinsically limited in their minimal period.
Fluorine excimer lasers currently offer the shortest commercially available laser wavelength of $\lambda_\mathrm{L} =157\,\mathrm{nm}$.  Even though shorter wavelengths will become available one day,
a natural limit is set by the size of the interfering particle. For instance, a rhinovirus of $10^6$\,amu has a diameter of 30\,nm, which would be comparable to 40\% of the grating period $d=\lambda_\mathrm{L}/2$ already at  $\lambda_\mathrm{L} =157\,\mathrm{nm}$.
And even for the densest metal clusters the density never exceeds  $2\times 10^4$\,kg/m$^3$. A gold cluster with  $10^{6}$\,amu thus measures already 5.4\,nm in diameter,  i.e. about 7\,\% of the $78\,$nm grating period produced by the mentioned excimer laser. Even though an optical grating would neither be clogged nor destroyed by an incident particle, the experienced effective potential will be smeared out if the particle size becomes comparable with or even larger than then grating period.

A further practical mass limit is given by the above mentioned collimation condition which can be expressed as a momentum condition. The transverse momentum must be smaller than the momentum kick imparted by the diffraction process:
\begin{equation}\label{MomentumTime}
    m< \frac{h}{d v_T} = \frac{h}{d v_L\Theta}   =  \frac{h}{d} \frac{t_1}{ D} =   \frac{h^2}{ 2d^2 k_B T     \Theta}
\end{equation}
Here $v_T$ and $v_L$ are the transverse and the longitudinal velocity, respectively, and  $t_1$ is the transit time in the collimation stage between $L_1$ and $L_{2}$.
This emphasizes the need for small transverse velocities $v_\mathrm{T}$, i.e. transverse cooling. While the cooling of atoms is an established laboratory technology, the cooling of clusters and molecules to below 1\,K is still a challenge.  A collimation to better than $v_T/v_L =$10\,$\mu$rad corresponds already to a ratio between the transverse and the longitudinal temperature of  $(v_T/v_L)^2 = 10^{-10}$.
In many current experiments, one therefore relies on selection rather than cooling.
According to Eq.\,(\ref{MomentumTime}) a reduction in $D$ will increase the mass limit, but at the expense of a reduction of the transmitted flux in proportion to  $D^2$. Also, increasing the flight time $t_1$ requires either a longer distance $L_1$ or a lower longitudinal velocity $v_\mathrm{L}$. But since the particle flux scales with  $L^2$ and $v_\mathrm{L}^2$ the diffraction of massive objects is bound to low signals.
At the right-hand side of Eq.\,\ref{MomentumTime} we replaced the longitudinal velocity by the most probable thermal speed. This equation leads to a mass limit of $10^6$ amu for a source temperature of 10\,K and a collimation to $\Theta\simeq 10\,\mu$rad, i.e. a transverse temperature of 1\,nK. This corresponds to the parameters of the gold cluster. A source of appropriate intensity at this temperature still has to be demonstrated.\\

So far, our discussion included geometrical and kinematic arguments as well as the filtering of molecules in the presence of van der Waals forces.  But also external or inertial forces can induce a fringe shift when they are oriented parallel to the grating vector. Their influence can be estimated using semiclassical arguments since, in the presence of conservative force fields, the shift of the beam envelope equals the displacement of the quantum interference pattern.

{\em Gravity} may cause a dispersive  fringe shift: In a horizontally oriented beam experiment the grating bars will ideally be oriented parallel to the line of gravity  $\textbf{g}$, however with an experimentally unavoidable misalignment angle $\epsilon_1$ . We require the gravitational shift of two contributing velocity classes $v$ and  $v+\Delta v$ to be smaller than the separation of two interference fringes. This leads to the requirement
\begin{equation}\label{VelocityWidth}
\Delta v/v \le v L_{2}h/mdgL^{2}\epsilon_1.
\end{equation}
For the gold clusters this yields $\Delta v/v \le 5\times10^{-7}/\epsilon_1$, implying that we require a collimation and maximal misalignment of better than  $\epsilon_1<10^{-5}$ for a practical experimental velocity bandwidth of $\Delta v/v= 5\%$.
It should be noted, however, that at a velocity of 1\,m/s and a flight distance of $L_1+L_2=2$\,m the gravitational free fall distance would already amount to 20\,m! This may still be feasible but it certainly poses a technological challenge.
The falling time would be shortened by a factor of ten if we used ten times faster gold clusters with a ten times better collimation, i.e. $\Theta=1\,\mu$rad.   The falling distance would thus be reduced by a factor of one hundred to merely 20\,cm.  The collimation requirement of $1\mu$rad appears to be a formidable task as well,  given our present-day technologies \cite{Hornberger2009}.

Also the {\em rotation of the Earth} shifts the interference patterns. We choose a coordinate system such that the angular frequency vector of the Earth is  $\Omega=\omega (0 , \cos \, \phi , \sin \,\phi)$, where $\omega=73\,\mu$rad/s and $\phi$ specifies our geographical latitude. The Coriolis acceleration is given by $ \bold{a}_C = 2\,\bold{v} \times \Omega $; if we orient the experiment vertically and such that $v_x=0$ and with grating bars aligned along $x$, the Coriolis acceleration will point along the slits and the contrast will only be reduced by angular misalignments.

In order to quantify this effect we set $v_x=v_y =\epsilon_2 \, v_L$ and  $v_z=v_L-gt$, where $\epsilon_2$ represents the angle between the molecular beam and gravity. We neglect the time dependence of $\textbf{v}$ due to the Coriolis force. A double integration of $\bold{a}_c$ over time yields the fringe shift.  Here, we are only interested in the displacement along the grating vector, i.e.
 \begin{equation}\label{CoriolisShift}
    y_c=-2\omega (v_L t^2 \epsilon_2 \sin \phi /2+(v_L t^2/2-gt^3/3) \epsilon_3 \cos \phi).
 \end{equation}
While the first term describes the Coriolis shift along $y$, the second term accounts for the finite alignment of the grating bars in relation to $x$, where $\epsilon_3$ measures the angle between the grating bars and the x-direction.
We require that the fringe shifts for different velocity classes should be smaller than one interference fringe, i.e.  $y_c\ll hH/(dm v_L)$. Assuming a flight time of  $t=v_L/g-(v_L^2/g^2-2H/g)^{1/2}$ and the derivative with respect to $v_L$ leads us to the velocity selection criterion
\begin{equation}\label{CoriolisShift2}
    \frac{\Delta v}{v} \le \frac{h H}{m v_L^2 d} \left[ \frac{v_L + gt}{v_L - gt} \epsilon_2 \sin \phi +\epsilon_3 \cos \phi \right]^{-1}.
 \end{equation}
With $H=1$\,m, $d=100$\,nm, $\phi=48^{\circ}$, $v_L=4.5$\,m/s and $M \simeq 10^{6}$\,amu,
we find that $\Delta v_L/v_L\le 1/ (4.1\times 10^2 \epsilon_2+36 \epsilon_3)$. This shows that the Coriolis force can be neglected even for a thermal molecular beam when the setup is aligned to better than $10^{-3}$\,rad.

To fulfill the combination of all requirements, i.e. collimation, velocity selection, orientation/alignment, high detection efficiency and source brilliance, for highly massive clusters is still a substantial challenge.  We therefore
proceed by illustrating an intermediate experiment in the mass range of around $m=30.000$  amu.  Assuming the possibility of sublimation at a temperature of 600\,K the de Broglie wavelength would still reach $\lambda_{\rm dB}=0.7$\,pm at a most probable velocity of 18\,m/s. When collimating the beam to $4\,\mu$rad on the screen by two slits with a width of $D=4\,\mu$m separated by a distance $L_1=1$\,m it should still be possible to identify two neighboring diffraction orders, separated by  the diffraction angle $\theta=\lambda_{\mathrm{dB}}/d=7\,\mu$rad.
The transverse coherence width then reaches 175\,nm at the location of the second collimator, where the $d=100$\,nm diffraction grating is placed; this is sufficient for a genuine double slit experiment. We choose the distances $L_2=L_1=1$\,m  and find a total transit time of $t_{\rm tot}=2t_1=110$\,ms, corresponding to a falling distance in the gravitational field of $H=6$\,cm.

A molecular flux of $\Phi =N L_1^2 v / D^2 Y^2 \eta \tau \Delta v =1.04\times 10^{16}$cm$^{-2}$s$^{-1}$sterad$^{-1}$ would be required to finally detect $N=1000$ individual molecules on the screen. Here we assume the source and collimator slits to be $Y=100\mu$m high, and take the accumulation time $\tau$ and the grating transmission $\eta$ to be $\tau=3600$\,s  and $\eta=1/3$, respectively. Starting from a thermal velocity distribution, the beam intensity will further be reduced by the required velocity selection, i.e. by about $\Delta v/v\simeq 5\%$.

\section{Poisson's spot}
The problem of small de Broglie wavelengths and low source intensities can often be alleviated in near-field diffraction experiments, where beam coherence and geometrical requirements are usually less demanding than in the far-field. Near-field effects comprise various phenomena, from diffraction at an edge~\cite{Gahler1991},  a grating~\cite{Chapman1995a,Nowak1998} or a circular obstacle~\cite{Reisinger2009} --up to the Talbot-Lau interferometry in an arrangement of two or three gratings~\cite{Brezger2002,Patorski1983,Clauser1994,McMorran2009}.
Here we focus on the diffraction pattern behind a radially symmetric obstacle of radius $R$, such as a disc or a sphere, that is illuminated by a point-like wave source \cite{Harvey1984a,Lucke2006}. The most prominent feature here is the appearance of a bright spot (Poisson's spot) in the center of the shadow region behind the obstacle, which is related to wave-like diffraction at the obstacle boundaries.

At a first glance it appears appealing to use this effect to demonstrate the wave-particle duality for very massive particles, as the mere existence of intensity in the dark field could be interpreted  as an indicator of the particle's wave nature. A second glance reveals, however, that the dispersive interaction between the particles and the obstacle walls must be taken into account. In particular, the presence of van der Waals forces can significantly obscure the spot even for neutral particles when they have a large polarizability. At the same time the attraction to the obstacle walls alone may already also explain the appearance of a bright spot in the dark-field, even if we take polarizable clusters to behave like billiard balls following classical Newtonian mechanics.
Our following theoretical treatment transcends earlier methods for near-field Poisson patterns~\cite{Dauger1996a} in that it now includes, for the first time quantitatively, the attractive interaction.

The geometry of the setup is sketched in Figure~\ref{Fig1}b. The beam is directed along the $z$-axis, with the points $z=0$, $z=L_1$, and $z=L_1+L_2$ defining the $xy$-planes of the source, the obstacle, and the detection screen, respectively. A circular pinhole of radius $R_0$ represents the source which emits a beam of particles with a collimation angle $\Theta$ towards an opaque obstacle of radius $R$ at the distance $L_1$. We are interested in the spatial density $w(\vu)$ of the particles another distance $L_2$ further downstream, as a function of the dimensionless screen coordinate $\vu = (x/R,y/R)$. In the paraxial approximation the size of both the source and the obstacle are taken small compared to the distances, $R_0,R \ll L_1,L_2$, and the beam is well collimated, $\Theta \ll 1$. The diffraction pattern of a monochromatic particle beam with the de Broglie wavelength $\lambda_{\rm dB} = h /m v_z$ is then given by
\begin{equation}
 w(\vu) = \frac{1}{\pi} \int_{|\vu_0| \leq 1} {\rm d}^2 \vu_0 \, \left| \psi \left( \left| \vu + \frac{L_2}{L_1} \frac{R_0}{R} \vu_0 \right| \right) \right|^2 . \label{eqn:poissonpattern}
\end{equation}
with $\psi$ defined in (\ref{eqn:psi1}), as follows from a phase-space description similar to ~\cite{Hornberger2004,Nimmrichter2008}. It is normalized to the constant density of particles on the screen that would be observed in the absence of an obstacle, $w(|\vu| \gg 1) = 1$.
We introduce the dimensionless parameters $\ell = (L_2+L_1)/L_1$ and $k = R^2 / (L_2 \lambda_{\rm dB})$, the main quantities characterizing the dimensionless amplitude function
\begin{equation}
 \psi (u) = \int_1^{\infty} {\rm d} s \, 2\pi k \ell s \, \exp \left( i \pi k \ell s^2 + i \phi (s) \right) \, J_0 \left( 2\pi k u s\right) \label{eqn:psi1}.
\end{equation}
It contains the Bessel function of the first kind $J_0$ and a phase $\phi(s)$ related to the interaction between the particles and the obstacle (see below). In the absence of the van der Waals interaction the diffraction pattern
associated with a point source at the origin $\vu_0 = (0,0)$ reads  $w_{p} (\vu) = |\psi (u)|^2$ \cite{Harvey1984a}. The integral in the amplitude function (\ref{eqn:psi1}) can be evaluated numerically by exploiting the exact result for $\phi=0$ when the lower integral bound is extended to zero.

\subsection{The ideal Poisson spot}
\begin{figure}
\centering
  \includegraphics[width=\textwidth]{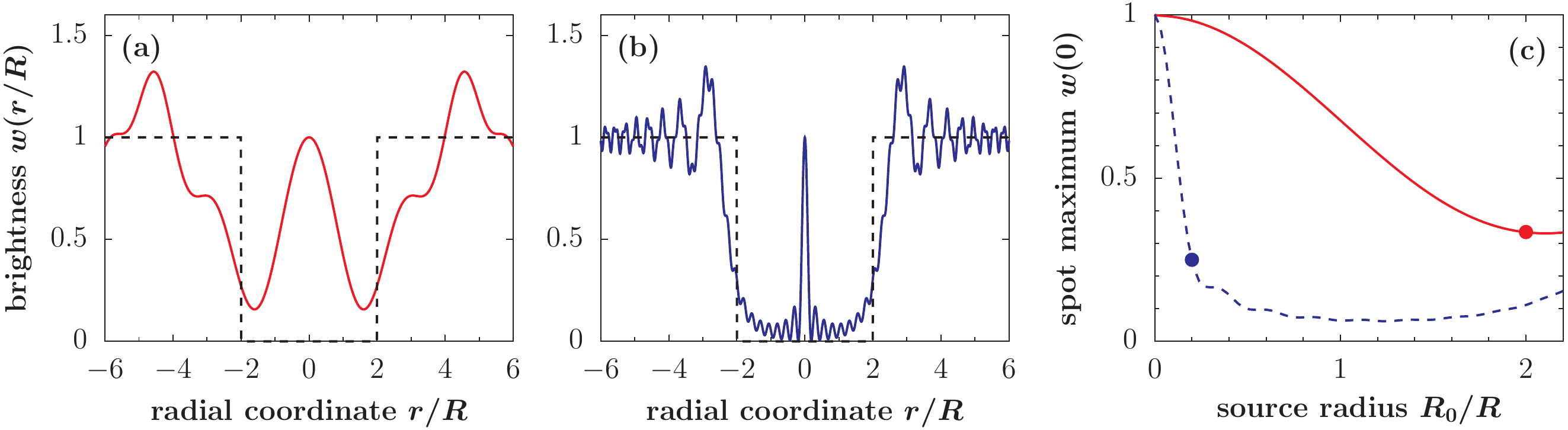}
  \caption{Figures (a) and (b) depict a radial cut of the Poisson diffraction pattern on the screen, assuming  an ideal point source and the absence of any van der Waals forces. The parameters $\ell= (L_2+L_1)/L_1$ and $k=R^2 / (L_2 \lambda_{\rm dB})$ are chosen as $\ell=2$ for both cases and as $k=0.2$ and $k=2$ for (a) and (b), respectively. The dashed line represents the classical shadow profile. The screen coordinate is given in units of the obstacle radius; the signal is normalized to the constant intensity in the absence of an obstacle. In (c), where a finite  extension of the source is taken into account, the height of the Poisson spot at the origin is plotted as a function of the source radius (in units of the obstacle radius). The cases (a) and (b) are here represented by the solid and by the dashed line, respectively. The visibility condition for the source radius (\ref{eqn:sourcecond}) is marked by a filled dot in both cases.
  } \label{fig:poissonideal}
\end{figure}

In Fig.~\ref{fig:poissonideal}(a) and Fig.~\ref{fig:poissonideal}(b), we plot the radial profile of the rotationally symmetric interference pattern (\ref{eqn:poissonpattern}) for  $k=0.2$ and $k=2$, respectively,  for a symmetric experimental  setup, i.e. $\ell = 2$.  The dotted line depicts the expected classical shadow profile which would be observed in the absence of diffraction. Its shadow region has a radius of $R_{\rm cl} = \ell R$ on the screen. The diffraction pattern exhibits a wavelike behavior with wide interference fringes for small $k$, while it approaches the classical shadow profile for large $k$, i.e. in the 'classical limit' $\lambda_{\rm dB} \to 0$ at finite distances from the symmetry axis. However, one can always observe an intensity peak at $\vu=0$ in the center of the classical shadow region. The spot is as bright as the classically expected intensity outside the shadow region, $w_p(0)=1$, independently of the setup geometry and the de Broglie wavelength. It is only the width of the spot that depends on both the wavelength and the geometry. The radius $R_s$ of the central spot is determined by the first zero of the Bessel function $J_0$ in (\ref{eqn:psi1}), which yields $R_s \approx 0.4 R / k = 0.4 L_2 \lambda_{\rm dB} / R$. Consequently, the successful observation of the spot requires only that its width be smaller than the width of the classical shadow  $R_{\rm cl} = \ell R$, which results in the rather lax condition $k\ell \gtrsim 0.4$. The condition is met exactly in Fig.~\ref{fig:poissonideal}(a), and by a factor of $10$ in~\ref{fig:poissonideal}(b). In a setup with an obstacle radius of $R=500\,$nm and $L_1=L_2=12.5\,$cm, the two plotted cases cover a range of de Broglie wavelengths $\lambda_{\rm dB} $ of $1\,{\rm pm}$ to $10\,$pm.

At first glance one might therefore think that no velocity selection is needed. This advantage, however, must be put into perspective as we have assumed an idealized point source, where the spot maximum is always $w(0)=1$, regardless of the incident de Broglie wavelength. In a real physical situation, the source has a finite extension and the spot is washed out because of the averaging over the source aperture, as described by Equation (\ref{eqn:poissonpattern}). This renders the height of the central spot wavelength-dependent. The reduction of the spot maximum $w(0)$ as a function of the source radius $R_0$ is plotted in Figure~\ref{fig:poissonideal}(c) for the cases of \ref{fig:poissonideal}(a) (solid line) and \ref{fig:poissonideal}(b) (dashed line). In the latter case, where the wave length is five times smaller, the spot vanishes more rapidly because it is narrower than in the former case. The spot starts to get lost in the background as soon as it is averaged over more than its width $R_s$ on the screen. Plugging this into Equation (\ref{eqn:poissonpattern}) we can estimate a condition for the source radius at which a pronounced Poisson spot can still be observed \cite{Lucke2006},
\begin{equation}
 R_0 \lesssim 0.4 \frac{L_1}{L_2} \frac{R}{k} =  0.4 \frac{L_1 \lambda_{\rm dB}}{R}. \label{eqn:sourcecond}
\end{equation}
The values given by this estimate are marked by full dots in Fig.~\ref{fig:poissonideal}(c). One notes from Eq.~(\ref{eqn:sourcecond}) that a larger distance $L_1$ between source and obstacle allows for larger source extensions. However, the particle beam intensity decreases quadratically with $L_1$, which limits the possible source distance in practice. The distance to the screen $L_2$, on the other hand, determines the width of the ideal Poisson spot through $k$. In the plotted examples we set $L_1 = L_2 = 12.5\,$cm. For $k=0.2$ (corresponding to $\lambda_{\rm dB} = 10\,$pm) one still obtains a pronounced central peak for a source pinhole radius of $R_0= R = 500\,$nm, as demonstrated by the solid line in Fig.~\ref{fig:poissonideal}(c). At the same time, the spot is strongly smeared out for $k=2$ (dashed line, corresponding to $\lambda_{\rm dB} = 1\,$pm).

Although the condition (\ref{eqn:sourcecond}) resembles the limit for the collimation slit aperture $D$ of a far-field interferometer, whith the obstacle radius $R$ playing the role of the grating period $d$, the collimation requirements for a Poisson spot experiment are much less stringent than in the far-field case. The diffraction pattern (\ref{eqn:poissonpattern}) is in fact \textit{independent} of the collimation angle $\Theta$, provided that it lies within the range $R/L_1 \ll \Theta \ll 1$. The much stricter collimation requirement $\Theta < \lambda_{\rm dB}/d$ of the far-field setup is therefore relaxed in practice, which is an advantage of the Poisson spot scheme, as has been demonstrated in \cite{Reisinger2009}. In case of a realistic particle beam with a finite longitudinal coherence, one should keep in mind that the diffraction pattern (\ref{eqn:poissonpattern}) must also be averaged over the distribution of de Broglie wavelengths $\lambda_{\rm dB}$.

Ultimately, the admissible pinhole radius of the source is bounded by $R_0 < R (L_1+L_2)/L_2$ to ensure that there is a shadow region at all, where the Poisson spot may emerge. This observation of particles in a classically forbidden area on the screen would thus be a clear indication of matter wave diffraction, provided we could neglect all particle-wall interactions.

\subsection{Influence of the particle-obstacle interaction}
A complete discussion of the mass limitations of Poisson's spot has to include the effect of the dispersive interaction between the diffracted particles and the obstacle's surface. In fact, the ideal spot pattern discussed so far can only be observed with light, fast and weakly polarizable particles such as atoms and D$_2$ molecules \cite{Reisinger2009}. The interaction potential $V((x^2+y^2)^{1/2},z)$ between a highly polarizable nanoparticle and the radially symmetric obstacle, however, is not negligible anymore. In our case of a well collimated beam and a small obstacle dimension we can account for it by introducing the eikonal phase term $\phi(r) = - \int {\rm d} z \, V(r,z) / \hbar v_z$ \cite{Nimmrichter2008}, which modulates the diffraction pattern through the amplitude function (\ref{eqn:psi1}). The interaction is here approximated by a Casimir-Polder-type attractive potential which diverges at the obstacle wall. As a consequence, the obstacle is effectively enlarged from $R$ to $R(1+\eta)$ because particles passing the obstacle at a distance smaller than $  \eta R$ will hit the wall and be adsorbed. A good estimate for  $\eta$ is obtained from the minimal classical impact parameter that still yields an asymptotically outgoing trajectory of a particle impinging upon the obstacle plane parallel to the $z$-axis \cite{Landau1960,Nimmrichter2008}.

We start by considering the diffraction at a nanosphere since such obstacles can be fabricated with a surface smoothness on the atomic level~\cite{BritishBiocell2010}. Taking a metallic sphere of radius $R$ in the range of a few hundred nanometers, positioned at $z=L_1$, we approximate the interaction by the asymptotic Casimir-Polder potential with an infinite wall \cite{mostepanenko-casimir} spanned by the tangential plane on the surface of the sphere
\begin{equation}\label{Casimir}
V(x,y,z) = -C_4 / ((x^2+y^2+(z-L_1)^2)^{1/2}-R)^4
\end{equation}
Here, the Casimir parameter $C_4$ is the same as used in the far-field discussion of Sect.~2.
While the computation of the exact Casimir-Polder potential in this geometry  requires advanced numerical treatments \cite{PhysRevA.81.030502},
our approximation (\ref{Casimir}) is conservative since it overestimates the interaction strength at distances $\ge R$ from the sphere surface---as the sphere bends away from the particle trajectory.

 \begin{figure}
\centering
  \includegraphics[width=\textwidth]{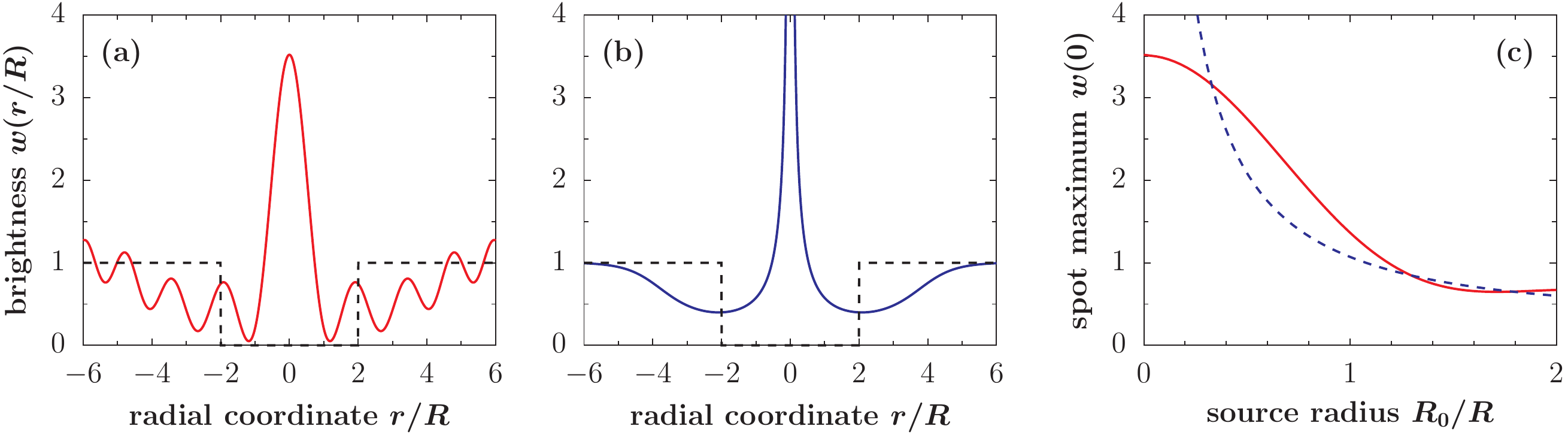} \\
  \vspace{2mm}
  \includegraphics[width=\textwidth]{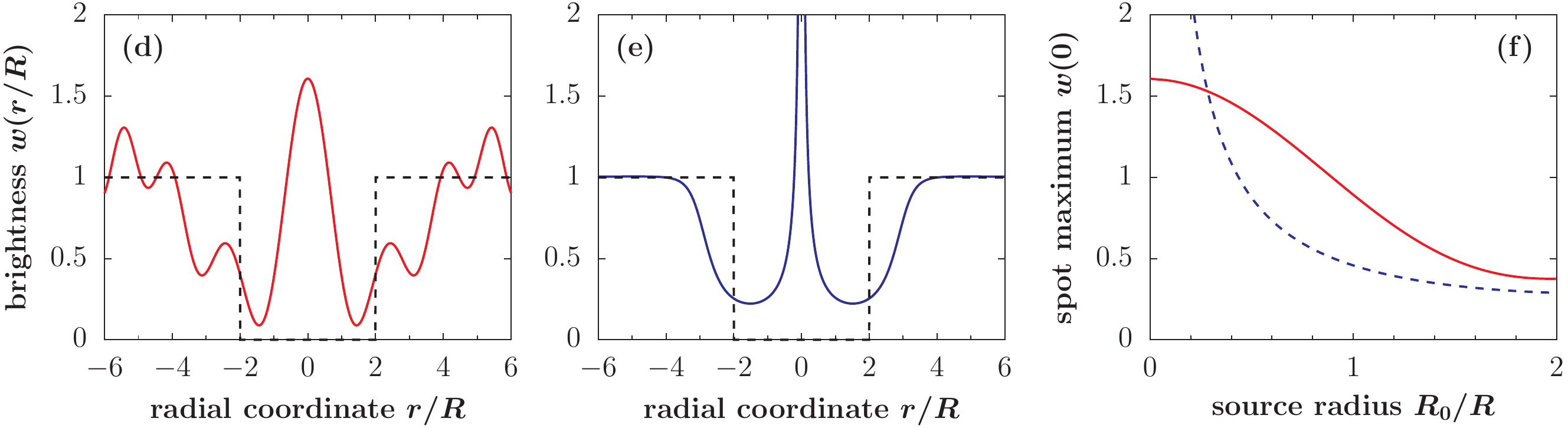}
  \caption{(a) Radial profile of the Poisson diffraction pattern caused by a point source with the same geometry parameters $k=0.2$ and $\ell=2$ as in Fig.~\ref{fig:poissonideal}(a). The interaction between Au$_{100}$ gold clusters and the sphere of size $R=500\,$nm is taken into account as described in the text. (b) Corresponding radial intensity profile assuming a classical description of the particle motion. It diverges at the origin. The dashed line represents the ideal shadow projection of the obstacle in both cases. (c) Plot of the height of the central spot as a function of the finite source radius (in units of the obstacle radius). The quantum case (a) and the classical case (b) are represented by the solid line and by the dashed line, respectively.  The second panel (d)--(f) shows the equivalent plots if the obstacle sphere is replaced by a disc of the same radius and a thickness of $b=10\,$nm, which reduces the interaction strength significantly.} \label{fig:poissonphase}
\end{figure}

Figure \ref{fig:poissonphase}(a) depicts the Poisson spot of a point source if the dispersion interaction is taken into account. The geometry parameters $k=0.2$ and $\ell=2$ are the same as in the ideal case of Fig.~\ref{fig:poissonideal}(a), but we now apply our potential model for a spherical obstacle attracting and diffracting a Au$_{100}$ cluster with a mass of $m=19700\,$amu and a polarizability of $\alpha = 500\,$\AA$^3$. The dashed line in the plot marks the shadow projection of the obstacle. We note that the maximum of the Poisson spot clearly increases with growing attraction, when compared to the ideal case.

What might look like a benefit at first glance, is relativized by plot
 (b) where we show the intensity distribution predicted by a classical deflection model. It was computed using the classical analogue of the eikonal phase approximation \cite{Hornberger2004,Nimmrichter2008}. The deflection is modeled by an instantaneous and radially inward directed momentum kick $q(r) = - \int {\rm d} z \, \partial_r V(r,z) / v_z = \hbar \partial_r \phi(r)$. The plots show that for highly polarizable particles a purely classical reasoning suffices to explain a spot-like intensity accumulation in the center of the screen. In fact, the depicted classical intensity distribution caused by a point source exhibits a $1/u$-like divergence, which is related to the radial symmetry and is compensated by the area element $u {\rm d} u$ in the course of any integration over a finite area around the origin.

The assessment of such a Poisson spot experiment therefore requires a careful quantitative analysis of the measured intensity on the screen in order to rule out a classical alternative for the expected experimental data. This problem is aggravated with growing particle size and polarizability. In addition, the classical and the quantum description become even less distinguishable with growing source radius. This is demonstrated in part (c) of Fig.~\ref{fig:poissonphase}, where the height of the central spot in the quantum (solid line) and in the classical (dashed line) case is shown as a function of the source radius in units of the obstacle radius. In such a regime the observation of a central spot can no longer be used as an indicator for the quantum wave nature.
Instead of restricting the analysis to the central spot, one could alternatively take the outer intensity oscillations as a quantum signature in an experiment. These are, however, less pronounced and more easily averaged out over both a finite velocity distribution of the particles and an extended source.
Even if the quantum and the classical case coincide, the setup may still serve as a van-der-Waals lens for particle beams.

For the plots (d)--(f) in the lower panel of Figure \ref{fig:poissonphase} the spherical obstacle has been replaced by a metallic disc of thickness $b=10\,$nm. The interaction is now approximated by the Casimir-Polder potential $V(r,z) = -C_4 / (r-R)^4$ of an infinite plane acting on the particle during the time of flight $b/v_z$ past the disc. Since the accumulated phase is now smaller than in the case of a spherical obstacle, a lower Poisson spot is found in (d) compared to (a).
With such a thin disc obstacle the quantum interference effect is clearly distinguishable from the classical deflection model, as demonstrated in (f). Even for a realistic source radius $R_0=R=500\,$nm the quantum spot visibility exceeds its classical counterpart significantly.

For even heavier particles than the Au$_{100}$ clusters discussed here, as well as for larger $k$-values, the classical deflection of the particle trajectories may no longer be approximated by an instantaneous momentum kick. Also in the quantum case the eikonal approximation ceases to be valid and must be replaced by a more complicated semiclassical scattering transformation \cite{Nimmrichter2008}.

Finally we note that, apart from being altered by the dispersion force close to a surface, the quality of the Poisson spot is also influenced by the surface roughness of the obstacle. In the deuterium diffraction experiment reported in \cite{Reisinger2009} the surface roughness of the disc dominated the influence of the interaction potential and led to a significant diminuition of the spot. However, modern microfabrication techniques as well as the strong interaction of the large nanoparticles considered here allow us to neglect the effects of surface corrugations. In the setting of Fig.~\ref{fig:poissonphase}, where Au$_{100}$ clusters with a velocity of $v_z = 2.0\,$m/s are diffracted at a sphere (or disc) of radius $R=500\,$nm, the effective enlargement of the obstacle radius by the cut-off (particle capture) distance is as large as $\eta R = 39\,$nm, or $17\,$nm in case of the disc. This exceeds by far the surface roughness of spheres or discs whose corrugations can nowadays be kept on the Angstrom level ~\cite{BritishBiocell2010}.

\section{Conclusions}
In summary, our discussion shows that a straightforward extrapolation of conceptually simple ideas such as far-field diffraction at a grating or the observation of Poisson's spot behind a spherical or disc-shaped obstacle leads to non-trivial experimental challenges and may require a careful assessment of what can be observed.
A number of dephasing agents, such as gravity, the rotation of the earth and -- more than anything else -- the influence of particle-wall interactions have a strong and usually contrast-limiting influence.
Our analysis shows, however, also that present-day experiments are still far from any fundamental limit.
Although other experimental arrangements may be better adapted for pushing the ultimate mass and complexity limits of matter wave interferometry~\cite{Reiger2006,Gerlich2007}, we still envisage many interesting experiments in far-field diffraction and in observing Poisson's spot with large clusters and molecules, in particular also with the foreseeable advent of new nanofabrication techniques for ultra-thin diffractive elements.

\begin{acknowledgements}
We acknowledge support through the ESF program EuroQuasar MIME and the FWF project Z149-N16. MA and KH acknowledge fruitful discussions with Bodil Holst, University of Bergen.
\end{acknowledgements}



\end{document}